\documentclass{cimento}

\usepackage{graphicx}  

\title{Opening the path to hard X-/soft gamma-ray focussing: the ASTENA-pathfinder mission}
\shorttitle{The ASTENA-pathfinder mission}

\author{E.~Virgilli\from{ins:1}\ETC,
L.~Amati\from{ins:1},
N.~Auricchio\from{ins:1},
E.~Caroli\from{ins:1},
F.~Fuschino\from{ins:1},
M.~Orlandini\from{ins:1},
J.~Stephen\from{ins:1},
L.~Ferro\from{ins:2},
F.~Frontera\from{ins:2},
M.~Moita\from{ins:2},
P.~Rosati\from{ins:2},
M.~Caselle\from{ins:3}
\atque
C.~Ferrari\from{ins:4}}

\instlist{
\inst{ins:1} INAF/OAS Bologna, Italy
\inst{ins:2} Dipartimento di Fisica e Scienze della Terra, Universit\`a di Ferrara, Italy
\inst{ins:3} Karlsruher Institut für Technologie, Karlsruhe, Germany
\inst{ins:4} CNR/IMEM Parma, Italy and INAF/OAB Brera, Italy
\inst{ins:5} INAF/IASF Palermo, Italy
}

\shortauthor{E.~Virgilli et al.}

\begin{document}

\maketitle

\vspace{-20pt}
\begin{abstract}
Hard X-/soft gamma-ray astronomy is a crucial field for transient, 
nuclear and multimessenger astrophysics. However, 
the spatial localization, imaging capabilities and 
sensitivity of the measurements are strongly limited for the 
energy range $>$70~keV. To overcome these limitations, we have 
proposed a mission concept, ASTENA, submitted to ESA 
for its program “Voyage 2050”. We will report on a pathfinder 
of ASTENA, that we intend to propose to ASI as an 
Italian mission with international participation. It will be 
based on one of the two instruments aboard ASTENA: a 
Laue lens with 20~m focal length, able to focus hard 
X-rays in the 50-700~keV passband into a  3-d position sensitive 
focal plane spectrometer. The combination 
of the focussing properties of the lens and of the 
localization properties of the detector will provide 
unparalleled imaging and spectroscopic capabilities, thus 
enabling studies of phenomena such as gamma-ray bursts 
afterglows, supernova explosions, positron annihilation 
lines and many more.
\end{abstract}

\vspace{-24pt}
\section{Rationale and mission configuration}
High energy astrophysics has still unanswered questions 
mainly due to the poor sensitivity of the present 
instrumentation: transient events like gamma-ray 
bursts and in particular their X-ray 
afterglow emission, blazars and 
magnetars spectra are just a few examples of 
science that would benefit from higher sensitive
instruments. An exhaustive review 
of the science that can be tackled with future focusing 
optics can be found in~\cite{Frontera2021, Guidorzi2021}.
A substantial increase in sensitivity can be achieved 
by enabling hard x-ray focusing. At the moment, the 
only technology that would enable the focalization 
of hard X-rays is through Laue lenses which are based on 
diffractive crystals. Unfortunately, due to 
the outstanding accuracy required for optics alignment 
and the non-trivial requirements for an efficient focal 
plane soft gamma-rays detector, such a technology has not yet
been used in space. 
At present, new technologies 
for the production of effective optics and 
new materials for gamma-ray detection 
make the use of Laue's lenses more mature 
and feasible.
A hard x-ray mission concept named 
ASTENA (Advanced Surveyor of Transient Events and for 
Nuclear Astrophysics) has been proposed for 
focusing photons in the 50 -- 700~keV energy range.
The ASTENA mission consists of two complementary 
instruments: a Wide Field Monitor with Imaging and Spectroscopic 
capabilities (WFM-IS), based on the same technology of the THESEUS/XGIS 
instrument~\cite{amati22} and a Narrow Field Telescope (NFT). 
The NFT is based on a 
Laue lens made from bent crystals of silicon and germanium. The 
lens has a diameter of 3~m and a focal length (FL) of 20~m.  
As a precursor for the ASTENA mission, we propose an experiment 
based on NFT alone whose total weight would be suitable for the 
size of a possible light national mission, in response to the 
next upcoming ASI call. 
This configuration, called ASTENA-pathfinder, will allow us
to tackle a relevant scientific case with a low mission weight 
($<$ 200~kg). 

\begin{figure}
\centering
\includegraphics[width=4.2cm,keepaspectratio]{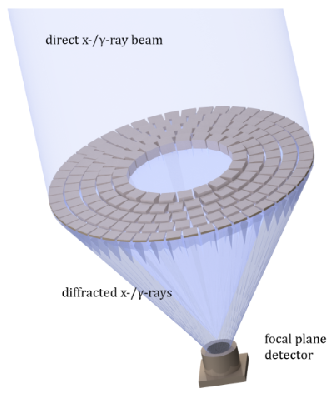}
\includegraphics[width=6cm,keepaspectratio]{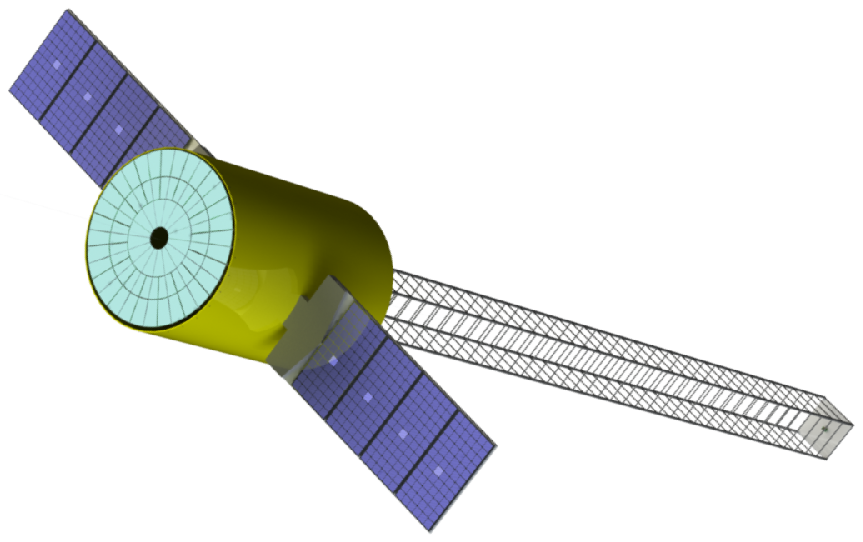}  
\caption{Left: Drawing of the Laue lens concept. Right: Drawing of the ASTENA 
pathfinder. The blue circular area represents the Laue lens, which 
is divided into independent modules. A position-sensitive detector 
is located at a focal distance of 20~m through an extendable mast.}
\vspace{-0.7cm}
\label{astenapathfinder}
\end{figure}

\section{Technical Development Activities}

Technical Development Activities
have been conducted in recent years in order 
to increase the maturity of a space instrument for 
focusing gamma-rays. The following sections describe 
such development activities separately for the Laue 
concentrator and the focal plane detector.

\begin{figure}[!h]
\centering
\includegraphics[width=5cm,keepaspectratio]{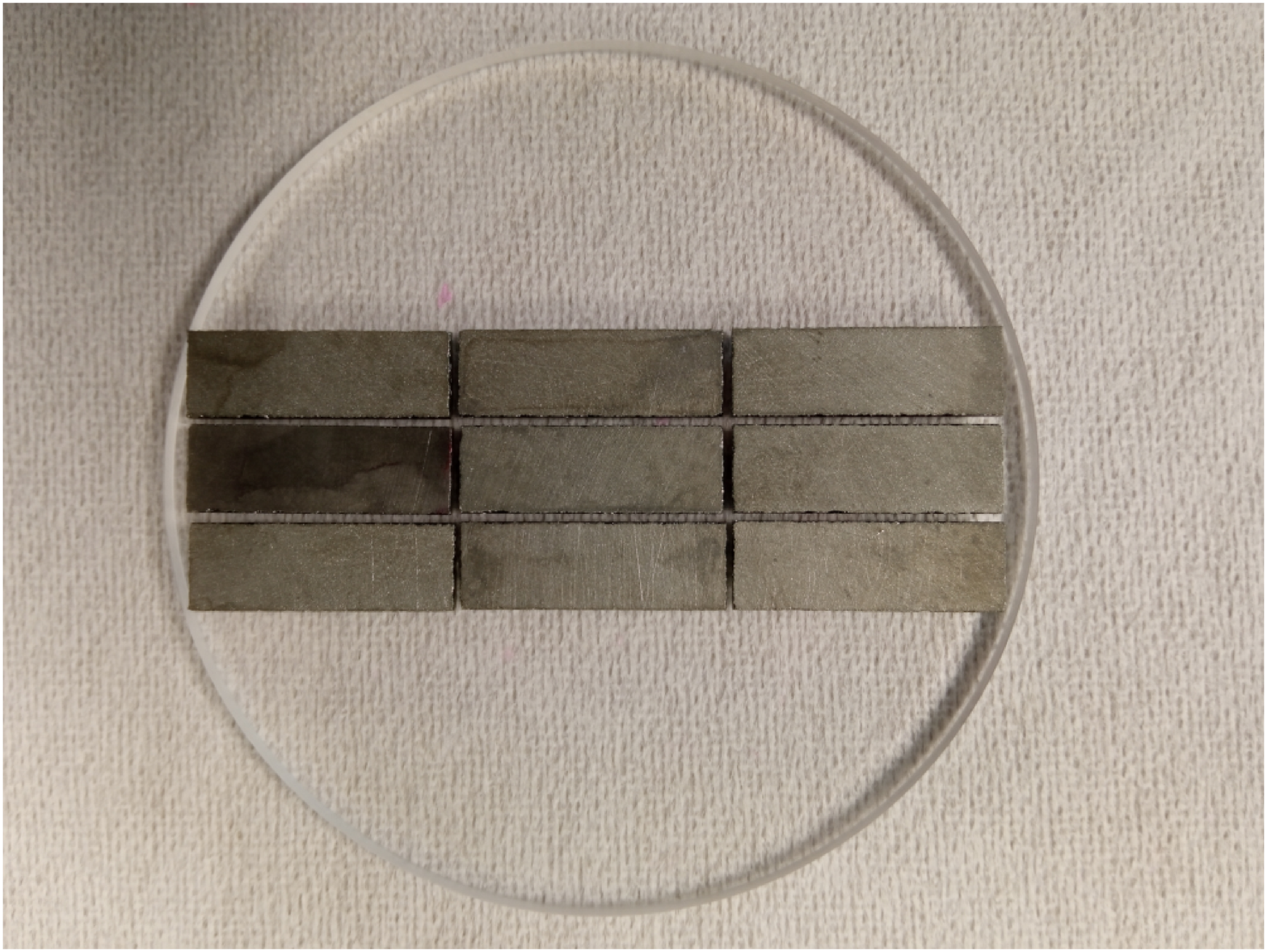}
\includegraphics[width=5.5cm,keepaspectratio]{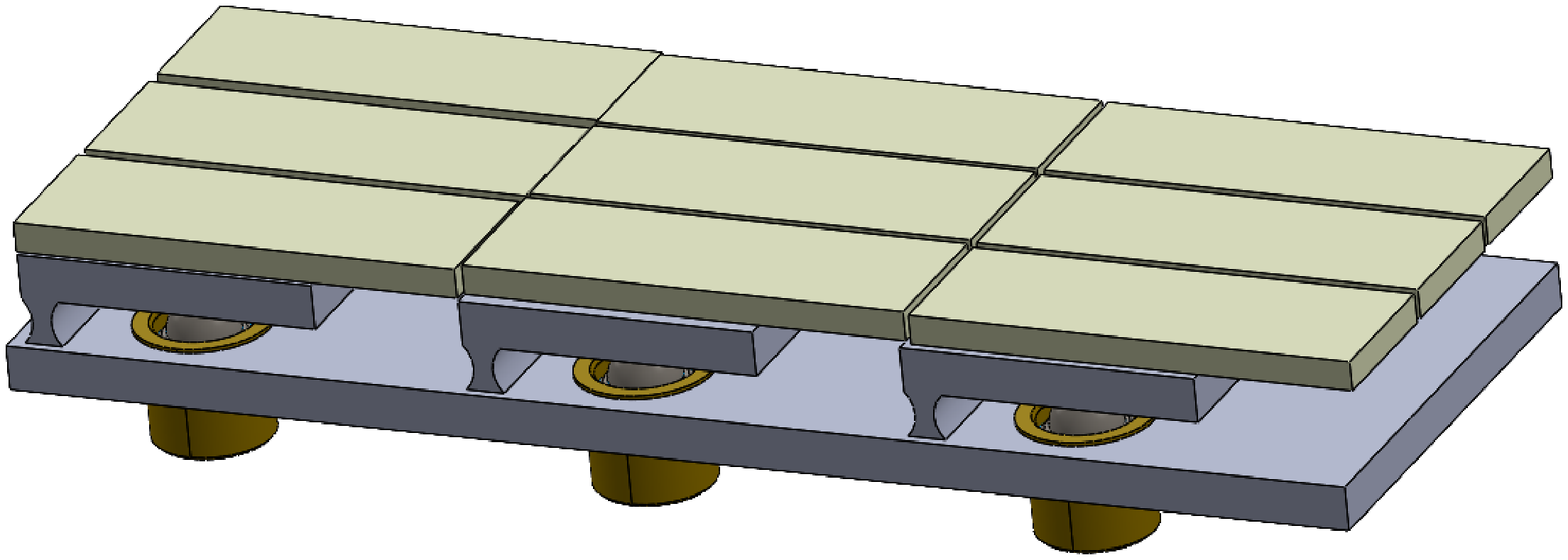}
\caption{Left: prototype of Laue lenses realized at 
Karlsruher Institut für Technologie (KIT) made with 9 
bent Germanium crystals mounted on a 4~mm quartz substrate 
with a bonding method adapted from the 
ball-bonding/ flip-chip technique largely used in 
microelectronics. Right: drawing of a 9 elements 
prototype which exploits the lamellae adjustable 
through micrometric screws.}
\vspace{-10pt}
\label{alignment}
\end{figure}

\begin{figure}
\centering
\includegraphics[width=10cm,keepaspectratio]{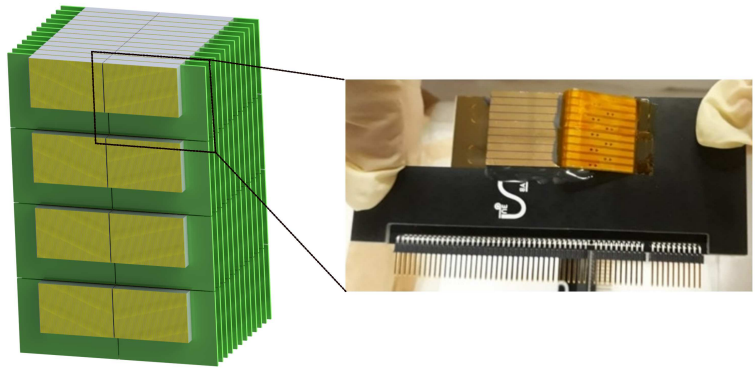}
\caption{Left: the configuration of a possible focal plane detector based 
on a stack of CZT crystals. Right: detail of one CZT detector developed 
within the 3DCaTM project, supported by ASI-INAF (adapted from 
\cite{caroli22}).}
\vspace{-0.7cm}
\label{3dcatm}
\end{figure}

\subsection{Developments of Laue lenses}
Laue lens technology is mainly limited by the large number of 
basic optical elements to be accurately prepared and 
properly oriented towards a common focal point. 
As these processes are currently performed manually, 
are prone to inaccuracy and time-consuming.
Our goal is to automate the process to reduce 
uncertainties and processing time.
Several R\&D activities have been made in the past 
for increasing the maturity of the Laue lenses.
Within the recent TRILL project~\cite{ferro22}, funded by 
ASI-INAF, CNR/IMEM (Parma) has optimised a method based 
on lapping one of the largest faces to provide the crystals 
the defined curvature. We have provided and qualified dozens 
of crystals by demonstrating the reproducibility of the 
crystal bending process at the nominal radius of 40~m 
(twice the FL). 
Various methods for positioning and bonding crystals 
were also investigated with the aim of finding a 
repeatable technique for a large number of 
crystals and with an accuracy 
of the order of a few arcsec.
The flip chip bonding technology, largely used 
in microelectronics, has been used in our specific
application at Karlsruher Institut für Technologie (KIT). 
This method allows to 
set at the same angle several crystals per minute with an 
accuracy better than 10~arcsec. However, it does not 
allow setting each crystal at its nominal angle  that 
also depends on the miscut angle with respect to the 
external surface, therefore the alignment procedure
relies also on the accuracy of the preparation of the 
sample. A method enabling bonding and fine alignment 
of the optical elements involves the use of elastically 
adjustable supports (called lamellae) operated by micro
screws (Fig.~\ref{alignment}-b). Each crystal is bonded 
through adhesive on the top of a lamella and fine threads micro screws 
ensures an alignment accuracy of the order of a few arcsec. A prototype 
with 9~crystals has been designed and will be realized and 
tested soon.

\subsection{Development of gamma-ray focal plane detector}
Technological activities are being performed by different groups in Europe 
\cite{kuvvetli14, caroli22} with the goal of developing 
semiconductor spectroscopic imagers suitable as focal plane for hard X-ray space telescopes.
In order to take advantage of a focusing instrument, requirements on the energy 
and spatial resolution, and on detection efficiency must be set. The requirements 
for such a detector are a high detection efficiency ($>$80\% @ 500~keV), in 
the energy range 10‐1000~keV and high performance spectroscopy ($\le$1\% FWHM @
511~keV). In addition, sub-millimeter spatial resolution in three dimensions ($\le$ 300 $\mu$m), 
fine timing resolution ($\le$1 $\mu$s) and polarimetry capability are needed. 
In this perspective, Cadmium Zinc Telluride (CZT)
have very high detection efficiency in the sub-MeV range even with only a 
few cm of materials, with good spectroscopic capability of about
1$\%$ FWHM @ 511~keV. They result to be particularly interesting as they provide 
high performance with the advantage of being usable at room temperature.
The baseline technology for these detectors is based on the segmentation of 
anodes and cathodes associated with a drift strip configuration for the 
anodic reading of the signals. 
 The use of anode and cathode strips allows to achieve 
a highly segmented sensor providing thousands of equivalent sub-millimetric 
voxels with a few tens of read-out channels (Fig.~\ref{3dcatm}) for each 
CZT sensor unit. In order to reach the required thickness, a stack of 
crystals is necessary.

\section{Conclusions}

We have presented an experiment called ASTENA-pathfinder based on a 
lightweight Laue lens made with curved Silicon and Germanium crystals. 
Thanks to the sensitivity achievable for both continuum 
 and nuclear line observations, the ASTENA-pathfinder will represent a 
 turning point for a number of still unanswered scientific questions.
ASTENA-pathfinder is a scientific and technological pioneer of the ASTENA 
mission concept which includes, in addition to the Laue lens of the NFT, a 
WFM-IS with unprecedented broad energy pass-band (2~keV - 20~MeV).
Current technological activities, supported by ASI and INAF, 
address both the complexity of building a Laue optics, that 
requires great precision for the alignment 
of thousands of basic components, and the realisation of a focal 
plane detector with fine 3-d segmentation, fast timing capabilities, 
high detection efficiency and high energy resolution.


\acknowledgments
This work is supported  by the ASI-INAF agreement N. 2017-14-H.O ”Studies 
for future scientific missions” and in particular within the Technological Readiness 
Improvement of Laue Lenses (TRILL) project and by the AHEAD-2020 Project grant agreement 
871158 of the European Union’s Horizon 2020 Programme.

\bibliography{virgilli-sif2022}

\begin{thebibliography}{1}
\expandafter\ifx\csname url\endcsname\relax\def\url#1{\texttt{#1}}\fi
\expandafter\ifx\csname urlprefix\endcsname\relax\def\urlprefix{URL }\fi

\bibitem{Frontera2021}
\NAME{Frontera F., Virgilli E., Guidorzi C., Rosati P., Diehl R. \etal},
  \IN{Experimental Astronomy}{51}{2021}{1175}.

\bibitem{Guidorzi2021}
\NAME{Guidorzi C., Frontera F., Ghirlanda G., Stratta G., Mundell C.~G.,
  Virgilli E. \etal}, \IN{Experimental Astronomy}{51}{2021}{1203}.

\bibitem{amati22}
\NAME{Amati L., Labanti C., Mereghetti S. \etal}, \TITLE{{The X/Gamma-ray
  Imaging Spectrometer (XGIS) for THESEUS and other mission opportunities}},
  presented at \TITLE{Space Telescopes and Instrumentation 2022: Ultraviolet to
  Gamma Ray}, Vol. 12181, International Society for Optics and Photonics (SPIE)
  2022.

\bibitem{caroli22}
\NAME{{Caroli} E., {Abbene} L., {Zappettini} A., {Auricchio} N. \etal},
  \IN{Mem.~Soc.~Astron.~Italiana}{93}{2022}{201}.

\bibitem{ferro22}
\NAME{{Ferro} L., {Virgilli} E., {Moita} M., {Frontera} F. \etal}, \TITLE{{The
  TRILL project: increasing the technological readiness of Laue lenses}},
  presented at \TITLE{Society of Photo-Optical Instrumentation Engineers (SPIE)
  Conference Series}, edited by \NAME{{den Herder} J.-W.~A., {Nikzad} S. \atque
  {Nakazawa} K.}, Vol. 12181 of \emph{Society of Photo-Optical Instrumentation
  Engineers (SPIE) Conference Series} 2022.

\bibitem{kuvvetli14}
\NAME{{Kuvvetli} I., {Budtz-J{\o}rgensen} C., {Zappettini} A., {Zambelli} N.,
  {Benassi} G., {Kalemci} E., {Caroli} E., {Stephen} J.~B. \atque {Auricchio}
  N.}, \TITLE{{A 3D CZT high resolution detector for x- and gamma-ray
  astronomy}}, presented at \TITLE{High Energy, Optical, and Infrared Detectors
  for Astronomy VI}, edited by \NAME{{Holland} A.~D. \atque {Beletic} J.}, Vol.
  9154 of \emph{Society of Photo-Optical Instrumentation Engineers (SPIE)
  Conference Series} 2014.

\end{thebibliography}
\bibliographystyle{varenna}

\end{document}